\documentclass[12pt]{article} 
\usepackage{amssymb}
\usepackage{graphicx}
\begin{document} 
\begin{center}

{\bf Ultraperipheral nuclear interactions}

\vspace{2mm}

I.M. Dremin\footnote{e-mail: dremin@lpi.ru}\\

{\it Lebedev Physical Institute, Moscow, Russia}

\end{center}

Keywords: proton, nucleus, ultraperipheral interactions, cross section,
form factor, impact parameter
                                             
\vspace{1mm}

\begin{abstract}
Large-distance ultraperipheral collisions of two relativistic ions
are considered. The clouds of photons surrounding the ions are responsible
for their distant electromagnetic interaction. The perturbative approach and 
the method of equivalent photons are described. It is shown that the total 
cross section of these collisions increases rapidly with increasing energy and 
is especially large for heavy ions. Some experimental data and their comparison 
with theoretical approaches are described. Further proposals are discussed.
\end{abstract}

\vspace{1mm}

PACS: 25.75.-q, 34.50.-s, 12.20.-m \\

\vspace{1mm}

Contents.

1. Introduction

2. The early history

3. Electron-positron pair production in ultraperipheral collisions
according to the  perturbation theory

   3.1. The differential distributions

   3.2. The bound-free processes

4. The equivalent photon approximation

5. The preasymptotical behavior of the cross sections

6. Theoretical analysis of exclusive dilepton production

7. Comparison with experimental data

8. Searches for new physics

9. Conclusions

Acknowledgements

\section{Introduction.}

Ultraperipheral nuclear collisions are distinguished from others
by the nature of the interacting fields. They happen when the ions move not
close enough to interact strongly. Then the electromagnetic fields surrounding 
the ions enter the game.
We concentrate here on the pure ultraperipheral collisions where the photons 
from the two electromagnetic clouds surrounding both ions collide\footnote{
Sometimes they are called as the two-photon processes. More photons can be
involved in the interaction (radiative corrections to the two-photon graphs).}. 
Interest in them is related to the fact that electromagnetic fields become 
extremely compressed in the longitudinal direction and very strong at high 
velocities of the ions. The cross section of these processes increases at high 
energies even more rapidly than the strong interaction cross section. 
This feature opens the way to studies of strong electromagnetic fields and 
their possible non-linear effects.

The peripherality of these interactions is characterized by
the transverse distance between the trajectories of the centers of two
colliding ions, called the impact parameter $b$. For ultraperipheral 
collisions $b$ must be larger than the
sum of ions' radii $b>R_1+R_2$. Otherwise the ions interact strongly.
The total cross sections of strong hadronic interactions at present energies 
are very large. The ion collisions with small impact parameters are studied,
e.g., for searches of some effects due to production of the quark-gluon plasma.
The mean multiplicities of particles created by strong interactions are very 
high. Therefore, the particles produced by ultraperipheral processes would
be lost in the huge background from strong interactions. The special selection
criteria dictated by the kinematics of ultraperipheral processes must be 
imposed to separate them.
At the impact parameters, slightly exceeding the sum of the radii, the 
exchanged photon can excite one of the ions interacting directly with quarks 
and producing some bosons. The strong interactions get partly involved. The 
theoretical treatment becomes more complicated. That is why we will not 
consider such processes called as the photoproduction (or photonuclear) 
reactions. 

The present review is rather brief. It ia aimed on those who just start to get
acquainted with this problem. Its main purpose is to be a guide
to the papers where the discussed problems are presented in detail.
Therefore, it concentrates on some particular aspects
of ultraperipheral collisions related to studies at colliders.
It deals mostly with works done during the last decade with some
references to the previous stages. To shorten the review, no Figures 
and graphs abundantly shown in many papers are
demonstrated but multiple references to them are given with short resume
of obtained conclusions.

We start with a brief reminder of the early history of the problem of
ultraperipheral nuclear collisions. The perturbative approach to its solution 
is described. The asymptotical energy behavior of the cross section of the
electron-positron pair production in ultraperipheral collisions
is demonstrated. The higher order corrections 
are discussed in connection with the preasymptotical energy dependence of
the cross section. The method of equivalent photons is formulated and applied
to the calculation of the cross sections. The special features of the bound-free 
processes, where the produced electron becomes bound to one of the ions, are 
considered. The nuclear form factors and the suppression mechanism are discussed.
Some experimental data are compared with theoretical predictions. The comparison
motivates further analysis of the main assumptions used in the theoretical 
approaches. Searches for new physics in ultraperipheral processes are described.

\section{The early history}

Almost a century ago, in 1924, Fermi \cite{fer1, fer2} considered the problem of 
interaction of charged objects with matter: "Let's calculate, first of all, 
the spectral distributions corresponding to those of the electric field 
created by a particle with electric charge, $e$, passing with velocity, $v$, 
at a minimum distance, $b$, from a point, $P$." He obtained the formula for 
the intensity of the electromagnetic field created in this process. It 
was used in 1934 by Weizs\"acker \cite{wei} and Williams \cite{wil} for their 
formulation of the method of equivalent photons as discussed below.

The same year, Landau and Lifshitz \cite{lali}, impressed by the prediction
of positrons in the Dirac-sea theory, used the Dirac equation and
calculated the asymptotical behavior of the cross section of the production
of the electron-positron pair in the electromagnetic fields of the colliding 
relativistic nuclei. It happened to increase very rapidly at high energies $E$
as $\ln ^3\gamma $ where $\gamma =E/M $ is the Lorentz-factor of the colliding 
ions of mass $M$. That was a test for the newly born Dirac theory for the
positron. It is remarkable that this paper \cite{lali} 
was published almost immediately after the discovery of positrons
in cosmic ray interactions in 1932 (published in 1933 \cite{ande}).

Three years later, Racah \cite{rac} got the expression for the cross section 
in lowest order perturbation theory (Born approximation) which contained some 
preasymptotical terms increasing slower than $\ln ^3\gamma$.

These papers gave the start for more detailed theoretical studies of such 
processes. Experimental research became extremely intensive after high energy 
colliders got into the operation.

The very careful and 
detailed review of theoretical predictions and some early experimental 
data was given in 1975 by the Novosibirsk group \cite{bgms}. Let us mention 
also some later review papers \cite{beba, baur, bekn, baht, rvx, pmhk}. 

\section{Electron-positron pair production in ultraperipheral collisions
according to the perturbation theory}

As told above, the process of the electron-positron pair production in
ultraperipheral interactions of ions was the first one described theoretically.
In these collisions, the~two colliding protons or nuclei interact 
electromagnetically but not hadronically. They effectively miss each other
with no change of their states. They
interact only by the photon clouds, which create the electron-positron
pairs. No nuclear transitions appear at small transferred momenta. 
The large spatial extention of electromagnetic fields and
their high intensity at large velocities lead to~
the strong energy increase (proportional to $\ln ^3\gamma $) of the cross 
section of these processes. The high density of the photon clouds
surrounding heavy ions is in charge of large coefficients in front of this law,
proportional to the squares of their electric charges $Z_1e$ and $Z_2e$. 
It is $Z^4$ times less for the proton (the hydrogen atom nucleus!)
collisions. These fields act only for a short time and the perturbation theory
is appropriate. The famous Racah formula \cite{rac} for the total cross 
section of the ultraperipheral production of the electron-positron pair in 
collisions of fast nuclei derived in the Born approximation looks like
\begin{equation}
\sigma _{Z_1Z_2\rightarrow Z_1Z_2e^+e^-}=\frac {28(Z_1Z_2\alpha ^2)^2}
{27\pi m_e^2}\left(l^3-6.36l^2+15.7l-13.8\right),
\label{rac1}
\end{equation}
where
\begin{equation}
l=\ln \frac {2(p_1p_2)}{M_1M_2}=\ln \frac {s_{nn}}{m^2}=\ln (4\gamma _c^2)
\label{ln}
\end{equation}
and $m_e$ is the electron mass, $m$ is the nucleon mass, $p_i$ are the 
4-momenta of colliding ions with masses $M_i$ (considered equal on the right 
hand side), $\gamma _c$ is their Lorentz-factor in the center-of-mass system,
$s_{nn}$ is the squared total energy per colliding nucleon pair. The formula 
contains the preasymptotical terms proportional to $l^2$ and $l$ slower 
increasing with energy increase. The small mass of the electron in the 
denominator favors large values of the cross section. The ultraperipheral 
production cross sections for heavy lepton pairs 
($\mu ^+\mu ^-$ or $\tau ^+\tau ^-$) can be obtained at relativistic energies 
from the Racah formula (\ref{rac1}), which does not take into account the
form factors of the colliding objects, by inserting their masses in place 
of the electron mass\footnote{The form factors are usually accounted in the
framework of the equivalent photon approximation.}.
The cross sections are proportional to the inverse squares of the lepton masses 
and, therefore, become much smaller than those for the electron pairs.

In terms of the energy per the pair of colliding nucleons $\sqrt {s_{nn}}$ 
the Racah formula (\ref{rac1}) can be rewritten as
\begin{equation}
\sigma _{Z_1Z_2\rightarrow Z_1Z_2e^+e^-}=\frac {28(Z_1Z_2\alpha ^2)^2}
{27\pi m_e^2}\left(\ln^3 \frac {s_{nn}}{8.3m^2}+2.2\ln \frac {s_{nn}}{8.3m^2}
+0.4\right).
\label{rac2}
\end{equation}
This formula absorbs the strongest correction terms $l^2$ in Eq. (\ref{rac1})
inside the leading term. The numerical factor in the argument of the
logarithms is responsible for doing that.
Therefore, this formula can be directly applied for studies
of the preasymptotical behavior of the cross section. This is important
in view of newly constructed NICA and FAIR facilities with energies
$\sqrt {s_{nn}}$ about 10 GeV. Surely, the leading term dominates 
at the colliders RHIC and LHC with available energies of hundreds and 
thousands GeV.

All terms of Eq. (\ref{rac2}) are positive at $\sqrt {s_{nn}}>$3 GeV and
the leading term dominates at $\sqrt {s_{nn}}>$6 GeV. These energies are
below those at NICA and FAIR. Thus the Racah formula predicts quite noticeable 
effect already at energies about 10 GeV. In particular, the values of the 
cross section for PbPb collisions
are 1.4 kb at $\sqrt {s_{nn}}$=10 GeV; 22.8 kb at 100 GeV and 97.5 kb at 1 TeV.
The small mass of electrons defines so large values of the cross sections.

This formula was confirmed also by considering the Feynman diagrams
with two photons emitted by colliding ions and producing the electron-positron
pair. That is why the ultraperipheral collisions are often called as 
the two-photon processes.
Correspondingly, the higher order corrections due to the additional photons
emitted by ions were evaluated. These graphs are reproduced in many publications
on this subject.
The estimated part of the Coulomb correction to the Racah formula proportional
to $l^2$ is negative \cite{iss, lms, geku, baz}:
\begin{equation}
\sigma _C =-\frac {56}{9\pi }\frac {Z^4\alpha ^4}{m_e^2}f(Z)l^2,
\label{coul}
\end{equation}
where
\begin{equation}
f(Z)=(Z\alpha )^2\sum _{n=1}^{\infty }\frac {1}{n(n^2+(Z\alpha )^2)}.
\label{fZ}
\end{equation}
It is negligibly small for protons but becomes essential for heavy ions. Its 
account for Pb ions leads to the replacement of the factor 8.3 in 
Eq. (\ref{rac2}) by 16 approximately. The preasymptotical behavior of the cross 
section changes. The above estimates at different energies should be
corrected correspondingly. At 10 GeV the cross section becomes more than twice
smaller. The estimates are not reliable enough because the term $l^3$ is not
dominating anymore. Even at the LHC energies the cross section becomes smaller 
by about 13$\%$. Energies of NICA and FAIR are close to the threshold.

The unitarity corrections accounting for the loops of the light-by-light
scattering in Feynman diagrams are small for the ultraperipheral graphs of
the electron-positron production. In distinction, the Coulomb correction
becomes much less for muon pairs production while the role of the unitarity
corrections increases \cite{hkse}. At the same time, these conclusions and
quantitative estimates of the cross section values according to
Eqs (\ref{rac2}) and (\ref{coul}) can
change with account of the nuclear form factors \cite{bz}. These problems are 
crucial in connection with the so called ultraperipherality parameter
introduced below. It is related to the numerical factors discussed above.

The multiple pair production was also estimated by different theoretical 
methods (see, e.g.,\cite{iss, lms, baht}). The multiple pair production 
happens to be even more active at small impact parameters than the creation 
of a single pair. However, the total cross section is not very sensitive to
small impact parameters. The main contribution comes from large impact 
parameters. Therefore the single pairs dominate. 

\subsection{The differential distributions}

The precise differential distributions of the electron-positron pairs produced
in the two-photon collisions are rather complicated. They contain 20 independent 
helicity amplitudes \cite{catu}. Matrix elements of the perturbative approach
squared become strongly intermixed in the differential distributions.
Some simplified expressions are written in the review \cite{bgms}. 
As an example, we show the leading term of the distribution 
of the mass $W$ of the produced $e^+e^-$ system (Eq. (5.27) in \cite{bgms}):
\begin{equation}
\frac {d\sigma }{dW^2}=\frac {2(Z_1Z_2\alpha )^2
\sigma _{\gamma \gamma \rightarrow e^+e^-}(W^2)}{3\pi ^2W^2}
\ln ^3\frac {(p_1p_2)}{M_1M_2}.
\label{W}
\end{equation}
The detailed studies of the characteristics  of the dilepton pairs production
is still at the very initial stage. Their 
analysis shows that the dominant contribution to the total cross section 
at relativistic energies is provided by the region of production of electrons
at small angles, small transverse momenta and small pseudorapidities of the 
pair\footnote{The problem of the widening of these distributions compared
to their expressions in the perturbative approach is considered in
the recent paper \cite{klmu}.}.
Therefore, the photons with small squared 4-momenta (virtualities) 
are most important. They can be considered as being almost real (massless).
Then the differential cross section can be approximated by a product of the 
total cross section of the $\gamma \gamma $-transition to the electron-positron 
pair and the differential fluxes of photons which appeared already in Fermi's 
papers \cite{fer1, fer2}. Herefrom the equivalent photon approximation (see 
Section 4) follows \cite{wei, wil}. The maximum photon energy
$\omega _{max}=\sqrt {s_{nn}}/mb$ increases at higher collision energies
and becomes smaller at large impact parameters.

The formulas (\ref{rac1}) and (\ref{W}) are valid for the point-like source of 
the electromagnetic radiation. The charge distribution inside the colliding
protons and heavy nuclei must be taken into account. The intensity of the 
photon fields
depends on the transverse distance (the impact parameter $b$) between the 
centers of the colliding nuclei. Therefore their radii $R_i$ enter the game
due to the requirement $b>R_1+R_2$. These problems are considered within 
the equivalent photon approximation. 

\subsection{Bound-free processes}

Before delving into these problems, let us mention the so called 
bound-free effect induced by the creation of the electron-positron pairs.
This name is used when the created electron is captured by one of the ions
while the positron flyes away. This is an important source of beam ions loss.
The charge-to-mass-ratio $Z/A$ changes and new ions do not follow the former 
trajectory. This loss puts some limits on luminosity. Such ions can damage the 
accelerator magnets, beam pipes and even its external safety walls at the
distances of hundreds meters. They deposit their energy in a localized region 
of the beam pipe and heat it. That is of practical importance for operation of 
accelerators in the heavy-ion modes. The capture cross section is higher
for more heavy ions. However, the cross section of these processes 
increases with energy increase only logarithmically \cite{beba, mhh}, 
i.e. slower than the main process which rises as the cube of the logarithm.
The total cross section of the ultraperipheral collisions of lead nuclei
can be as large as 200 kb at the LHC while the capture cross section is
about 200 b. The pair production with capture will become comparable with
the production of free pairs at lower energies of NICA and FAIR.
 
\section{The equivalent photon approximation}

The essense of the equivalent photon approximation is already demonstrated
by the Eq. (\ref{W}). The Feynman diagrams of all processes with two-photon 
interactions contain a blob describing the transformation
of these photons to some final states (e.g., $e^+e^-$ considered above).
Thus it can be described by the cross sections of these processes.
The missing element of the whole picture are the photon fluxes in between
the colliding charged objects. Namely they were the main purpose
of Fermi's research as clearly stated in the citation put at the beginning 
of this review. The photons carrying small fractions $x$ of the nucleon energy
dominate in these fluxes. The~distribution of equivalent photons generated by a moving
(point-like!) nucleus with the charge $Ze$ and carrying the small fraction of 
the nucleon energy $x$ integrated over the transverse momentum up to some value 
(see, e.g., \cite{blp}) leads according to the method of equivalent photons 
to the flux
\begin{equation}
\frac {dn}{dx}=\frac {2Z^2\alpha }{\pi x}\ln \frac {u(Z)}{x}.
\label{flux}
\end{equation}
The ultraperipherality parameter $u(Z)$ depends on the nature of colliding 
objects and created states. Its physical meaning is the ratio of the maximum 
adoptable transverse momentum to the nucleon mass as the only massless 
parameter of the problem. In the perturbative approach it should incorporate
the radiative corrections that change the preasymptotical dependence of the 
cross section (see comments after the Eq. (\ref{coul})). It differs numerically 
in various approaches 
\cite{bgms, bb1, kn, bkn, balt, kgsz, seng, zha, kmr, geom, vyzh}. 
It depends on considered processes, as well as on
charges $Z_ie$, sizes and impact parameters of colliding objects
(form factors and absorptive factors). The~impact parameters cannot be 
measured but, surely, should exceed the sum of the radii of colliding ions. 
Otherwise the strong (QCD) and photonuclear interactions enter the game. 
This~requirement can be restated as a bound on the exchanged transverse momenta,
such that the objects are not destroyed but slightly deflected by the collision
and no excitations or nuclear transitions happen.
The bound depends on their internal structure, i.e., on forces inside them.
These~forces are stronger for a proton than for heavy nuclei. Therefore protons
allow larger transverse momenta. The~quantitative estimates of the parameter
$u$ for different processes are obtained from comparison with experimental data
and confronted with theoretical approaches described in more detail in the next
section.

The equivalent photon approximation allows a clear separation into a purely
kinematical effect of photon fluxes and the dynamical cross sections of their
interactions. Besides the electron-positron pairs considered theoretically in 
Refs \cite{lali, rac} and observed, e.g., in Refs \cite{ada, abb}, other 
pairs of oppositely charged particles with even $C$-parity can be created in 
the two-photon 
collisions. For example, pairs of muons produced in ultraperipheral collisions 
are observed at LHC \cite{dynd, kha, at1, at2, at3}. The~light-by-light 
scattering described theoretically by the loop of charged particles is also 
detected at LHC \cite{atax, dent, aad}. Some neutral $C$-even bosons composed of 
quark-antiquark pairs can be produced in the two-photon interactions. This 
process is especially suitable for the compact theoretical
demonstration~\cite{geom} of the $\ln ^3\gamma$-law. 

The exclusive cross section of the production of the resonance $R$ 
in the two-photon collisions of nuclei $A$ can be written as
\begin{equation}
\sigma _{AA}(R)=\int dx_1dx_2\frac {dn}{dx_1}\frac {dn}{dx_2}
\sigma _{\gamma \gamma }(R),
\label{e2}
\end{equation}
where the fluxes $dn/dx_i$ for the colliding objects 1 and 2
are given by Equation (\ref{flux}) and (see Ref. \cite{bgms})
\begin{equation}
\sigma _{\gamma \gamma }(R)=\frac {8\pi ^2\Gamma _{tot}(R)}{m _R}
Br(R\rightarrow \gamma \gamma )Br_d(R)\delta (x_1x_2s_{nn}-m_R^2).
\label{e3}
\end{equation}
Here, $m_R$ is the mass of $R$, $\Gamma _{tot}(R)$ its total width and  
$Br_d(R)$ denotes the branching ratio to a considered channel of its decay.
$s_{nn}=(2m\gamma )^2, m$ is a nucleon mass.
The $\delta $-function approximation is used for resonances with small
widths compared to their masses. The resonance is registered according to the 
peak in the distribution of the effective mass of the decay products 
$\sqrt{x_1x_2s_{nn}}$. As you see, the perturbative matrix element
approach is replaced in the equivalent photon approximation by the
semiclassical probabilistic scheme accounting for the structure of Feynman 
diagrams.

The integrals in Eq. (\ref{e2}) can be easily calculated so that one gets
the analytical formula
\begin{equation}
\sigma _{AA}(R)=\frac {128}{3}Z^4\alpha ^2Br(R\rightarrow \gamma \gamma )Br_d(R)
\frac {\Gamma _{tot}(R)}{m_R^3}\ln ^3\frac {2um\gamma }{m_R}.
\label{e4}
\end{equation}
The asymptotical $\ln ^3\gamma $ behavior is valid again.
The factor $2mu/m_R=1/\gamma _0$ defines the preasymptotical behavior of the 
ultraperipheral cross section of production of the resonance $R$.
The structure of this formula is similar to that used for $e^+e^-$ production
(\ref{rac2}). The variations of the parameter $u$ can account for the
subleading terms proportional to $\ln ^2\gamma $.
The asymptotical limit is reached at 
\begin{equation}
\gamma \gg m_R/2um,
\label{asym}                 
\end{equation}
where the terms increasing slower than $\ln ^3\gamma $ can be neglected.

The parameter $u$ can be found from Eq. (\ref{e4}) if the exclusive cross
sections of the ultraperipheral production of $\pi ^0$-mesons or
parapositronium are measured. The analogous formulae are obtained \cite{kkss, bena}
for the creation of the $C$-odd states like $\rho ^0$-mesons or
orthopositronium.

\section{The preasymptotical behavior of the cross sections}

The rapid asymptotical increase of the total cross section of ultraperipheral
collisions as $\ln ^3\gamma $ poses the question about its comparison
with the total cross section of purely hadronic interactions which
can not increase stronger than $\ln ^2\gamma $ according to the Froissart 
theorem \cite{froi} stemming from general theoretical principles.
According to experimental data on proton-proton collisions their increase 
is even slower at present energies.

The cross section for a single neutral pion production in ultraperipheral
collisions of two protons according to (\ref{e4}) is compared in 
Ref. \cite{uoh} with experimental data at TeV energies on the corresponding 
production channel in strong interactions. This cross section is about 1 nb,
while single $\pi ^0$s are produced in strong interactions with cross sections
of the order of 0.3 mb. It is shown that the additional $\ln \gamma $ factor 
is not large enough and absolutely insufficient
for such ultraperipheral processes to dominate in proton-proton collisions
over strong forces at any realistic energies . The background for $\pi ^0$ 
production due to strong interactions at small impact parameters must be 
enormously large. Some special cuts should be imposed to separate the
ultraperipheral events. The specific kinematics of ultraperipheral processes 
can be used for such cuts as shown in Section 7.

The analogous estimates for collisions of heavy nuclei are more complicated
due to the lack of the information about the definite reaction channels.
One can just state that the large numerical factor $Z^4/A^{2/3}\approx 10^6$ 
in the ratio of the ultraperipheral to purely nuclear (strong) interactions 
would favour heavy nuclei compared to protons.

Another preasymptotical problem is related to the energy behavior of the
ultraperipheral cross sections in the lower energy region. The Racah formula 
written as (\ref{rac2}) clearly demonstrates the $\ln ^3\gamma $
asymptotics and shows that the numerical factor 8.3 determines the
preasymptotical behavior of the ultraperipheral cross section. Moreover
it is modified by the radiative corrections. This numerical 
factor transforms in the equivalent photon approximation to the 
ultraperipherality parameter $u(Z)$ which has a meaning of the ratio
of maximum adoptable transferred momentum to the nucleon mass. This parameter 
incorporates the form factors and the impact parameters suppression. It 
influences the estimates of the cross sections, especially at lower energies.
That becomes important, for example, at energies of NICA and FAIR.
The electron-positron pair (as well as para- and ortho-positronium)
production seems to be feasible there \cite{uoh}
if the optimistic results from studies at the LHC \cite{vyzh, agm, szcz} 
are taken into account and extrapolated. The $\pi ^0$-production asks for 
energies close to those of NICA and is very sensitive to the estimates
of the parameter $u$ shown in Section 7.

\section{Theoretical analysis of exclusive dilepton production}

The perturbative matrix element approach (\ref{rac1}), (\ref{coul}) clearly
demonstrates that the higher order corrections can change the preasymptotical
values of the calculated cross section by changing the numerical factor in 
the argument of the  $\ln ^3$-term. Its predictive power suffers also from the 
consideration of colliding objects in Feynman diagrams as the structureless 
point-like particles. This deficiency can be partially avoided within the 
equivalent photon approximation (\ref{e2}) by the insertion and the 
interpretation of the ultraperipherality parameter $u(Z)$.

To take into account the size and the electric charge distribution inside the 
colliding hadrons and ions one must somewhat generalize the Eq. (\ref{e2}). 

The density of the photon flux depends on the structure of colliding objects and
on their impact parameter\footnote{The flux in Eq. (\ref{flux}) is integrated 
over the transverse momentum and the impact parameter.}. In terms of the form 
factors $F$ of colliding objects it looks like \cite{baur, vyzh, dys}:
\begin{equation}
\frac {d^3n}{d^2bdx}=\frac {Z\alpha}{\pi ^2x}\left[ \int dq_{\perp }
q_{\perp }^2J_1(bq_{\perp })\frac {F(q_{\perp }^2+m^2x^2)}
{q_{\perp }^2+m^2x^2}\right]^2.
\label{form}                 
\end{equation}
The electromagnetic fields of relativistic ions look like the narrow
pancakes perpendicular to their trajectories and moving together with ions.
Their interaction can be approximated by the $\delta$-function in this plane
at small $x<0.1/mb$ \cite{bkn}.
It is strongest when the ions come close to one another.
However, the impact parameter between them should be larger than the sum 
of their radii. Otherwise they get involved in strong  
interactions, and huge background to the products of ultraperipheral 
interactions appears. To exclude such collisions, the cutoff factor
$P(\vert {\bf b}_1-{\bf b}_2\vert )$ is introduced. It is determined
by soft physics with low transferred momenta. Therefore it is non-perturbative
and phenomenological. If the colliding ions are considered as the black disks,
this factor forbids completely the impact parameters smaller than the sum
of ions radii and is written in the following way
\begin{equation}
P=\theta (\vert {\bf b}_1-{\bf b}_2\vert -R_1-R_2), 
\label{rig}
\end{equation}
where $R_i$ are their radii.

The Eq. (\ref{e2}) is generalized\footnote{The symbol $R$ must be replaced by 
$X=l^+l^-$ for dileptons. The cross section 
$\sigma_{\gamma \gamma \rightarrow l^+l^-}$ is given by the Breit-Wheeler
formula \cite{brwh}.} as
\begin{equation}
\sigma _{AA\rightarrow AAX}=\int dx_1dx_2d^2b_1d^2b_2\frac {d^3n}{d^2b_1dx_1}
\frac {d^3n}{d^2b_2dx_2}
\sigma _{\gamma \gamma \rightarrow X}P(\vert {\bf b}_1-{\bf b}_2\vert ),
\label{aax}
\end{equation}
The choice of the cutoff factor determines the ultraperipherality parameter 
$u(Z)$. If the heavy ions stay intact after the collision then the                                                                                                                                                                                                                                                                                                                                                                                                                                                                                                                                                                                                                                                                                                                                                                                                                                                                                                                                                                                      
Eq. (\ref{rig}) should be used. At smaller impact parameters they do not 
survive. The hope
that the form factors in Eq. (\ref{form}) satisfy this requirement 
automatically is hardly realistic. The photon flux computed in Ref. \cite{agm}
for PbPb interactions at the energy 5.02 GeV per nucleon pair (see Fig. 3a
in Ref. \cite{agm}) becomes much smaller if the additional cut according
to (\ref{rig}) is imposed on it even for the realistic form factor.
For processes with initial protons, the elastic scattering at small impact 
parameters can be taken into account. The spatial distribution of their
inelastic profile depends on the collision energy \cite{cul}. The cutoff
factor can be generalized either by the account of the proton opacity,
as proposed in \cite{hkr}, or by the Glauber modification of (\ref{rig}), 
as proposed in \cite{agm}. The suppression factor $S^2$ has been used
for the quantitative estimate of the cutoff factor on the cross section 
of the ultraperipheral processes:
\begin{equation}                           
S^2=\frac 
{\int \int d^2b_1d^2b_2 \frac {d^3n}{d^2b_1dx_1}\frac {d^3n}{d^2b_2dx_2}
P(\vert {\bf b}_1-{\bf b}_2\vert )}{\int _{b_1>0} \int _{b_2>0}
d^2b_1d^2b_2 \frac {d^3n}{d^2b_1dx_1}\frac {d^3n}{d^2b_2dx_2}}
\label{sq}
\end{equation}
Its evaluation depends on the choice of the lower limits of the impact
parameters in the numerator (e.g., compare Eq. (7) in \cite{dys} and
Appendix 3 in \cite{vyzh}). This is the main origin of disagreement
in the different choices of the ultraperipherality parameter mentioned above
\cite{bgms, bb1, kn, bkn, balt, kgsz, seng, zha, kmr, geom, vyzh}.

Let us note, however, that the formulas for the total cross sections considered 
above are useful for understanding the energy behavior of ultraperipheral
processes and some general estimates but they are not very practical in the direct 
applications to experimental results. The experimentally measured phase space 
volume is usually significantly smaller than the total one. The detector 
structure and possible backgrounds reduce it. The so-called fiducial cross 
sections taking into account these "cavities"  and the requirement of the
"ultraperipherality" are measured. The Monte Carlo generators, 
e.g., STARlight \cite{knsg} or SuperChic \cite{hlkr}, are often used both for 
the selection of the most favorable and admissible conditions and for the 
further comparison with experimental results. The corresponding cuts are 
imposed on the matrix elements or on the formulas, derived according to the
equivalent photon approximation, for the computation of the fiducial cross 
sections, i.e., those which account for experimental cuts. The advantage of 
the equivalent photon approximation compared to 
the perturbative calculations or the diagram approach is that the analogous 
calculations can be done analytically (up to computing some simple integrals). 
Thus one gets the possibility to compare different approaches and 
parametrizations with experimental data and 
control the accuracy of the equivalent photon approximation. These experimental
cuts were taken into account in the considered below papers 
\cite{vyzh, agm, szcz}.

The more general problem is related to the spatio-temporal inhomogeneities
of the considered electromagnetic fields. They can play a prominent role
in the production of secondary particles. In particular, it is shown in
\cite{apsh} that it can result in the increased number of soft photons, i.e.
in higher photon fluxes. Nonlinear effects of strong-field QED are related to
the string problem \cite{frts} and can become important in heavy-ion collisions.
In particular, the dynamically assisted Schwinger effect and the Franz-Keldysh
effect are considered in \cite{taya}. They could be studied with
a proper combination of ultraperipheral fields and laser fields.

\section{Comparison with experimental data}

Prediction of large cross sections of ultraperipheral heavy ion collisions
stimulated their experimental studies at RHIC \cite{ada} and 
LHC \cite{abb, dynd, kha, at1, at2, at3}. There are special signatures of
these processes. The dilepton pairs in the final
state have very small transverse momentum. There are two rapidity gaps that
separate the intact very forward ions from the dilepton pair. From the
theoretical side the main problem is the proper estimates of the 
photon fluxes, i.e. the evaluation of the parameter $u(Z)$.

Production of $e^+e^-$ pairs in heavy-ion collisions was studied first
at RHIC by STAR Collaboration \cite{ada} and then at LHC by ALICE
Collaboration \cite{abb}. The obtained rapidity and invariant mass distributions
for the exclusive $e^+e^-$ production by $\gamma \gamma $ interactions in
PbPb collisions at $\sqrt s$=2.76 TeV were compared in Refs \cite{agm, szcz} 
with theoretical results using the above formulas. Besides the rigid absorption 
factor of the black disks (\ref{rig}), its a'la Glauber model modification was
considered. What concerns form factors, three of them were put on the trial:
point-like, monopole and realistic which corresponds to the Fourier-transform 
of the Wood-Saxon charge-density distribution of the nucleus. The general
behavior of both distributions is well reproduced by theoretical results
except the region of small invariant masses below 2.3 GeV (see Fig. 5 in
\cite{agm}). The experimental cuts were imposed on the computed distributions.
It was concluded that the modification of the absorption factor 
is unimportant. Both precise and monopole form factors fit experimental results 
at high masses rather well within error bars while the point-like one declines 
from them.  The experimental distribution is higher than the theoretical one 
at small masses less than 2.3 GeV in \cite{agm} but declines slightly only
at that  single point in \cite{szcz} (see Fig. 3 there). The low masses become 
the most important ones at the lower energies of NICA and FAIR. This region 
must be carefully studied there.

Production of $\mu ^+\mu ^-$ pairs in pp collisions was first observed
in 1990 at CERN's Intersecting Storage Rings (ISR) \cite{antr}. However
the detailed experimental studies \cite{dynd, at1, at2, at3} and comparison with 
theory \cite{vyzh, agm, szcz} became possible only recently. 

The ATLAS data presented in Ref. \cite{dynd} are confronted to theoretical 
results in Refs \cite{agm, szcz} in the same manner as done above for $e^+e^-$ 
data. Rapidity and invariant mass distributions for the exclusive $\mu^+\mu ^-$
production by $\gamma \gamma $ interactions in PbPb collisions at $\sqrt s$=5.02
TeV are plotted (see Fig. 4 in \cite{agm} and Fig. 4 in \cite{szcz}). 
The agreement with theory is less satisfactory within the precision of 
experimental data than for electron-positron pairs, especially for the 
case of the point-like form factor, as expected.

The very detailed comparison with experiment is done in the paper \cite{vyzh}.
It helped to get more definite information on the parameter $u(Z)$.
The experimental results from pp collisions at 13 TeV \cite{at1} and
PbPb collisions at 5.02 TeV per nucleon pair \cite{at2} were considered.

The total cross section of the ultraperipheral production of muon pairs 
in the equivalent photon approximation is
\begin{equation}
\sigma (ZZ(\gamma \gamma )\rightarrow ZZ\mu ^+\mu ^-)=\frac {28}{27}
\frac {Z^4\alpha ^4}{\pi m^2_{\mu }}\ln^3\frac {u^2s_{nn}}{4m_{\mu }^2}.
\label{vz}
\end{equation}
Let us note that the energy dependence in Eqs. (\ref{e4}) and (\ref{vz}) is 
the same for $m_R=2m_{\mu }$ as expected. The~preasymptotical behavior is 
determined by the factor $u$.

The cuts on the invariant mass of the $\mu ^+\mu ^-$ pair (on the fraction 
$x$ in Eq. (\ref{flux})), on the muon transverse momentum (on the differential
$p_T$-distribution of $\gamma \gamma $ processes) and on the pseudorapidity 
(required by the detector geometry) were imposed on Eq. (\ref{aax}) both for
pp-collisions at 13 TeV ($Z$=1) and for PbPb-collisions at 5.02 TeV per
nucleon pair. The corresponding integrals are easily computed. These cuts 
reduce drastically the cross section values.

For example, for pp-processes the value 0.22 $\mu $b according to Eq. (\ref{vz})
is reduced to 3.35 pb. If corrected for absorptive effects~\cite{dys} it 
gives \mbox{3.06 $\pm$ 0.05 pb}. The~chosen cuts coincide with those imposed in 
studies of the ATLAS collaboration~\cite{at1} which measured the value 
3.12 $\pm $ 0.07 (stat.) $\pm $ 0.10 (syst.) pb. The~SuperChic Monte Carlo
program~\cite{hlkr} which incorporates, in principle, both ordinary and 
ultraperipheral processes predicts 3.45 $\pm $ 0.06 pb. Theoretical results are 
in agreement with experimental data and show that ultraperipheral processes 
dominate over other sources in this fiducial volume. Analogous conclusions
were obtained for lead-lead collisions \cite{vyzh}. Here, due to $Z^4$
enhancement the~measured fiducial cross sections are on the $\mu $b scale 
compared to pb's for pp-collisions.

The ultraperipherality parameter $u$ is the least precisely determined element 
of the equivalent
photon approximation. As described above, its evaluation crucially depends on 
two main factors accounting for the impact parameter suppression $P(b)$ and
for the charge distribution inside ions (form factors $F(q)$).
The~careful treatment of form factors of protons and nuclei with 
account of the photon virtuality (see also Refs \cite{bb1, bkn} where the
problem was treated in more detail) and the suppression factors \cite{vyzh} lead
to its values $u_{pp}\approx 0.2$ for pp and $u_{PbPb}\approx 0.02$ for
PbPb-collisions within the factors about 1.5 which depend on the particular
shape of the form factors. At the very beginning, these values of the parameters 
$u(Z)$ were qualitatively estimated from general physics arguments and then 
confirmed by the successful comparison of theoretical predictions with 
experimental data. 

It is remarkable that these values of the ultraperipherality parameter $u$
agree quite well with those obtained in Ref. \cite{bena} for the
ultraperipheral cross sections of $\pi ^0$-production according to Eq.
(\ref{e4}) at RHIC and LHC energies. Their values (28 mb at LHC) shown in
the Table of Ref. \cite{bena} easily lead to $u_{PbPb}\approx 0.013$.
This value agrees up to the factor 1.5 with that shown above.

The knowledge of the ultraperipherality parameters
helps to estimate possible effects at the lower energies of
NICA and FAIR. They favor the ultraperipheral processes of $e^+e^-$ and
positronia production \cite{uoh} there, while the $\pi ^0$-production
is questionable because the argument of the logarithm in Eq. (\ref{e4})
is close to 1 and, therefore, the non-leading terms similar to those
in Eq. (\ref{rac2}) must be taken into account
(available near the threshold at the estimates of \cite{vyzh} 
and unavailable at the estimates of \cite{bena})\footnote{The earlier estimates 
\cite{geom} based on the extremely strong cuts of the impact parameters imposed in 
\cite{kmr} produced the much higher threshold for such processes.}.

\section{Searches for new physics}

The clean channel of $\gamma \gamma $ interactions in ultraperipheral collisions 
is often discussed in connection with searches for new physics. Before the 
Higgs-boson discovery it was actively debated as one of the possible sources 
for its creation (see, e.g., \cite{vgs, pio}). Nowadays, the main topics include 
searches for supersymmetric particles 
\cite{vgs, owz, scpi, kmrs, hkss, khoz, harl, bliu, gnrvz}, magnetic monopoles
\cite{tho, abbo, moed}, gravitons and possible extra spatial dimensions of the 
Kaluza-Klein theory with large compact dimensions in addition to the 4 
dimensions of Minkowski spacetime \cite{pio, anp}, axion-like pseudoscalar 
particles \cite{knap, berd, bald, smx, coel, siki, inki}, which would  induce  
anomalous  scattering  of  light-by-light, radions \cite{rich}, 
unparticles \cite{caoz}, impact of supersymmetry on properties of the
observed particles (e.g., the virtual sparticles in Feynman diagrams for
the anomalous magnetic moment of the tau lepton \cite{beli}).
There are no experimental signatures of these effects yet. Some lower limits, 
e.g., on the masses of supersymmetric particles \cite{pdg} or on the masses of 
the spin 0, 1/2, or 1 Dirac monopoles \cite{abbo} are established.

The very detailed proposal of searches for the supersymmetric particles in
ultraperipheral proton-proton collisions at the LHC was presented recently in 
the Ref. \cite{gnrvz}. It is proposed to search for charginos\footnote{Two
charginos and four neutralinos are the supersymmetric partners of the
electroweak bosons.} with masses about 100 GeV. Even though the LHC results 
exclude within a large interval of theoretical parameters their production 
with masses below 1 TeV according to \cite{pdg}, this conclusion is not valid
in the case when the masses of lightest chargino and lightest neutralino are
approximately equal. Namely this rather exotic possibility with the lightest 
chargino somewhat heavier than the lightest neutralino is studied.
The leading order Feynman diagrams for chargino pairs production in 
ultraperipheral collisions look similar to those for production of the 
electron-positron pairs considered above. The estimates of the total cross 
sections give the values 2.84 fb for pp collisions at 13 TeV and 21.2 pb
for PbPb collisions at 5.02 TeV per nucleon. If both protons are required
to be registered in forward detectors\footnote{It allows strong background 
suppression and complete reconstruction of event kinematics.} and the 
experimental cuts similar to the described above cuts for muon pairs data
are imposed, the fiducial cross section in proton-proton
processes is reduced to 0.72 fb. The background due to muon pairs and 
pile-up is estimated. It is shown that the peaks from chargino and muons
are well separated and the chargino peak is clearly visible over the pile-up
if the special cut on the longitudinal momentum of the final state system
is used (see Fig. 6 in Ref. \cite{gnrvz}).

\section{Conclusions}

The ultraperipheral processes provide very important information about
strong {\bf electromagnetic} fields. Experimental data on dilepton production 
are successfully described by the equivalent photon approximation.
The interaction of high energy photons
in the dense electromagnetic clouds surrounding relativistic protons and 
heavy ions opens ways to studies of new physics in the processes of production 
of new objects. The theoretical methods of their description are well developed
and prove their applicability when compared with experimental data.

\vspace{6pt}

{\bf Acknowledgement}

This work was supported by the RFBR project 18-02-40131 and RAN-CERN program.


\begin{thebibliography}{99}
\bibitem{fer1}
Fermi E {\em Z. Physik} {\bf 29} 315 (1924) 
\bibitem{fer2}
Fermi E {\em Nuovo Cim.} {\bf 2} 143 (1925),  
translated from Italian by M. Gallinaro and S. White in arXiv:hep-th/0205086
\bibitem{wei}
Weizs\"{a}cker C F V {\em Z. Physik} {\bf 88} 612 (1934) 
\bibitem{wil}
Williams E J {\em Phys. Rev.} {\bf 45} 729 (1934) 
\bibitem{lali}
Landau L D, Lifshitz E M {\em Phys. Z. Sowjetunion} {\bf 6} 244 (1934) 
\bibitem{ande}
Anderson C D {\em Phys. Rev.} {\bf 43} 491 (1933)
\bibitem{rac}
Racah G {\em Nuovo Cimento} {\bf 14} 93 (1937)
\bibitem{bgms}
Budnev V M et al. {\em  Phys. Rep.} {\bf 15} 181 (1975) 
\bibitem{beba}
Bertulani C A, Baur G {\em Phys. Rep.} {\bf 163} 299 (1988) 
\bibitem{baur}
Baur G et al. {\em Phys. Rep.} {\bf 364} 359 (2002)
\bibitem{bekn}
Bertulani C A, Klein S K, Nystrand J {\em Ann. Rev. Nucl. Part. Sci.} {\bf 55}
271 (2005) 
\bibitem{baht}
Baur G, Hencken K, Trautmann D {\em Phys. Rep.} {\bf 453} 1 (2007) 
\bibitem{rvx}
Ruffini R, Vereshchagin G, Xue S {\em Phys. Rep.} {\bf 487} 1 (2010) 
\bibitem{pmhk}
Di Piazza A et al. {\em Rev. Mod. Phys.} {\bf 84} 1177 (2012) 
\bibitem{iss}
Ivanov D Y, Schiller A, Serbo V G {\em  Phys. Lett. B} {\bf 454} 155 (1999)
\bibitem{lms}
Lee R N, Milstein A I, Serbo V G {\em Phys. Rev. A} {\bf 65} 022102 (2002)
\bibitem{geku}
Gevorkyan S R, Kuraev E A {\em J. Phys. G} {\bf 29} 1227 (2003)
\bibitem{baz}
Baltz A J  {\em Phys. Rev. C} {\bf 71} 024901 (2005)
\bibitem{hkse}
Hencken K, Kuraev E A, Serbo V G {\em Phys. Rev. C} {\bf 75} 034903 (2007)
\bibitem{bz}
Baltz A J {\em Phys. Rev. C} {\bf 80} 034901 (2009)
\bibitem{catu}
Carlson C E, Tung W K {\em Phys. Rev. D} {\bf 6} 147 (1972)
\bibitem{klmu}
Klein S R et al. arXiv:2003.02947
\bibitem{mhh}
Meier H et al. {\em Phys. Rev. A} {\bf 63} 032713 (2001) 
\bibitem{blp}
Berestetsky V B, Lifshitz E M, Pitaevsky L P
\emph{Kvantovaya Electrodinamika};
Fizmatlit: Moscow, Russia, 2001.
\bibitem{bb1}
Baur G, Bertulani C A {\em Z. Phys. A} {\bf 330} 77 (1988)
\bibitem{kn}
Klein S R, Nystrand J {\em Phys. Rev. Lett.} {\bf 84} 2330 (2000)
\bibitem{bkn}
Guclu et al. {\em Annals Phys.} {\bf 272} 7 (1999)
\bibitem{balt}
Baltz A J et al. {\em Phys. Rev. C} {\bf 80} 044902 (2009)
\bibitem{kgsz}
Klusek-Gawenda M, Szczyrek A {\em Phys. Lett. B} {\bf 763} 416 (2016)
\bibitem{seng}
Sengul et al. {\em Eur. Phys. J. C} {\bf 76} 428 (2016)
\bibitem{zha}
Zha et al. {\em Phys. Lett. B} {\bf 789} 238 (2019)
\bibitem{kmr}
Khoze V A, Martin A D, Ryskin M G arXiv:1902.08136
\bibitem{geom}
Dremin I M {\em  Int. J. Mod. Phys. A} {\bf 34} 1950068 (2019)
\bibitem{vyzh}
Vysotsky M I, Zhemchugov E V {\em Phys. Usp.} {\bf 189} 975 (2019)
\bibitem{ada}
Adams J et al. [STAR Collab.] {\em Phys. Rev. C} {\bf 70} 031902 (2004) 
\bibitem{abb}
Abbas E et al. [ALICE Collab.] {\em Eur. Phys. J C} {\bf 73} 2617 (2013) 
\bibitem{dynd}
Dyndal M [ATLAS Collab.] {\em Nucl. Phys. A} {\bf 967} 281 (2017)
\bibitem{kha}
Khachatryan V et al [CMS Collab.] {\em Phys. Lett. B} {\bf 772} 489 (2017) 
\bibitem{at1}
Aaboud M et al. [ATLAS Collab.] {\em  Phys. Lett. B} {\bf 777} 303 (2018) 
\bibitem{at2}
Aaboud M et al. [ATLAS Collab.] ATLAS-CONF-2016-025. 2016.
\bibitem{at3}
Arratia M arXiv:1611.05145
\bibitem{atax}
Aaboud M et al. [ATLAS Collab.] {\em  Nat. Phys.} {\bf 13} 852 (2017) 
\bibitem{dent}
d'Enterria D et al. [CMS Collab.] {\em Nucl. Phys. A} {\bf 982} 791 (2019)
\bibitem{aad}
Aad G et al. [ATLAS Collab.] {\em Phys. Rev. Lett.} {\bf 123} 052001 (2019)
\bibitem{kkss}
Kotkin G L et al. {\em Phys. Rev. C} {\bf 59} 2734 (1999)
\bibitem{bena}
Bertulani C A, Navarra F {\em Nucl. Phys. A} {\bf 703} 861 (2002)
\bibitem{froi}
Froissart M {\em Phys. Rev.} {\bf 123} 1253 (1961) 
\bibitem{uoh}
Dremin I M {\em Universe} {\bf 6} 4 (2020) 
\bibitem{klei}
Klein S R {\em Phys. Rev. C} {\bf 97} 054903 (2018)
\bibitem{agm}
Azevedo C, Goncalves V F, Moreira B D {\em Eur. Phys. J. C} {\bf 79} 432 (2019)
\bibitem{szcz}
Szczurek A arXiv:1810.06249
\bibitem{brwh}
Breit G, Wheeler J A {\em Phys. Rev.} {\bf 46} 1087 (1934)
\bibitem{dys}
Dyndal M, Schoeffel L {\em Phys. Lett. B} {\bf 741} 66 (2015) 
\bibitem{cul}
Dremin I M {\em Physics} {\bf 1} 33 (2019)
\bibitem{hkr}
Harland-Lang L A, Khoze V A, Ryskin M G {\em JHEP} {\bf 03} 182 (2016)
\bibitem{knsg}
Klein S R et al. {\em Comput. Phys. Commun.} {\bf 212} 258 (2017) 
\bibitem{hlkr}
Harland-Lang L A, Khoze V A, Ryskin M G {\em Eur. Phys. J. C} {\bf 79} 39 (2019)
\bibitem{apsh}
Aleksandrov I A, Plunien G, Shabaev V M {\em Phys. Rev. D} {\bf 100} 116003 (2019) 
\bibitem{frts}
Fradkin E S, Tseytlin A A {\em Phys. Lett. B} {\bf 163} 123 (1985) 
\bibitem{taya}
Taya H {\em Phys. Rev. D} {\bf 99} 056006 (2019)
\bibitem{antr}
Antreasyan D et al, Preprint CERN-EP/80-82 (1990)
\bibitem{vgs}
Vidovic M, Greiner M, Soff G {\em Phys. Rev. C} {\bf 47} 2288 (1993) 
\bibitem{pio}
Piotrzkowski K {\em Phys. Rev. D} {\bf 63} 071502 (2001) 
\bibitem{owz}
Ohnemus J, Walsh T F, Zerwas P M {\em Phys. Lett. B} {\bf 328} 369 (1994)
\bibitem{scpi}
Schul N, Piotrzkowski K {\em Nucl. Phys. Proc. Suppl.} {\bf 179} 289 (2008)
\bibitem{kmrs}
Khoze V A et al. {\em Eur. Phys. J. C} {\bf 68} 125 (2010)
\bibitem{hkss}
Harland-Lang L A et al. {\em Eur. Phys. J. C} {\bf 72} 1969 (2012)
\bibitem{khoz}
Khoze V A, Martin A D, Ryskin M G {\em J. Phys. G} {\bf 44} 055002 (2017)
\bibitem{harl}
Harland-Lang L A et al. {\em JHEP} {\bf 1904} 010 (2019)
\bibitem{bliu}
Beresford L, Liu J {\em Phys. Rev. Lett.} {\bf 123} 141801 (2019)
\bibitem{gnrvz} 
Godunov S I et al.  {\em JHEP} {\bf 01} 143 (2020)
\bibitem{tho}
't Hooft G {\em Nucl. Phys. B} {\bf 79} 276 (1974)
\bibitem{abbo}
Abbott B. et al. [D0 Collab.] {\em Phys. Rev. Lett.} {\bf 81} 524 (1998) 
\bibitem{moed}
Acharya B et al. {\em Phys. Lett. B} {\bf 782} 510 (2018)
\bibitem{anp}
Abers S C, Norbury J W, Poyser W J {\em Phys. Rev. D} {\bf 62} 116001 (2000) 
\bibitem{knap}
Knapen S et al. {\em Phys. Rev. Lett.} {\bf 118} 171801 (2017)
\bibitem{berd}
Bruce R et al. arXiv:1812.07688 
\bibitem{bald}
Baldenegro C et al.  {\em Phys. Lett. B} {\bf 795} 339 (2019)
\bibitem{smx}
Shaken S, Marsh D E, Xue S arXiv:2002.06023
\bibitem{coel} 
Coelho C et al. arXiv:2002.06027
\bibitem{siki}
Sikivie P arXiv:2003.02206
\bibitem{inki}
Inan S C, Kisselev A V arXiv:2003.01978
\bibitem{rich} 
Richard F arXiv:1712.06410
\bibitem{caoz} 
Cakir O, Ozansoy K O {\em Eur. Phys. J. C} {\bf 56} 279 (2008)
\bibitem{beli}
Beresford L, Liu J arXiv:1908.05180
\bibitem{pdg}
Tanabashi M et al. [Particle Data Group] 
{\em Phys. Rev. D} {\bf 98} 030001 (2018) 

\end{thebibliography}
\end{document}